\documentclass[10pt,conference]{IEEEtran}\usepackage[]{graphicx}\usepackage[usenames,dvipsnames]{color}
\makeatletter
\def\maxwidth{ %
  \ifdim\Gin@nat@width>\linewidth
    \linewidth
  \else
    \Gin@nat@width
  \fi
}
\makeatother

\definecolor{fgcolor}{rgb}{0.345, 0.345, 0.345}

\usepackage{framed}
\makeatletter
 {\par\unskip\endMakeFramed%
 \at@end@of@kframe}
\makeatother

\definecolor{shadecolor}{rgb}{.97, .97, .97}
\definecolor{messagecolor}{rgb}{0, 0, 0}
\definecolor{warningcolor}{rgb}{1, 0, 1}
\definecolor{errorcolor}{rgb}{1, 0, 0}
\newenvironment{knitrout}{}{} 

\usepackage{alltt}

\usepackage{balance}
\usepackage[square,numbers,sort]{natbib}
\usepackage{amsmath, amssymb, amsfonts}
\usepackage{algorithmic}
\usepackage{graphicx}
\usepackage{textcomp}
\usepackage{xcolor}
\usepackage[hyphens]{url}
\IfFileExists{upquote.sty}{\usepackage{upquote}}{}
\begin{document}

\title{Underproduction: An Approach for Measuring Risk in Open Source Software}

\author{\IEEEauthorblockN{Kaylea Champion}\IEEEauthorblockA{University of Washington\\Email: kaylea@uw.edu}\and\IEEEauthorblockN{Benjamin Mako Hill}\IEEEauthorblockA{University of Washington\\Email: makohill@uw.edu}}

\maketitle
 
\thispagestyle{plain}
\pagestyle{plain}

\begin{abstract}
The widespread adoption of Free/Libre and Open Source Software (FLOSS) means that the ongoing maintenance of many widely used software components relies on the collaborative effort of volunteers who set their own priorities and choose their own tasks. We argue that this has created a new form of risk that we call `underproduction' which occurs when the supply of software engineering labor becomes out of alignment with the demand of people who rely on the software produced. We present a conceptual framework for identifying relative underproduction in software as well as a statistical method for applying our framework to a comprehensive dataset from the Debian GNU/Linux distribution that includes 21,902 source packages and the full history of 461,656 bugs. We draw on this application to present two experiments: (1) a demonstration of how our technique can be used to identify at-risk software packages in a large FLOSS repository and (2) a validation of these results using an alternate indicator of package risk. Our analysis demonstrates both the utility of our approach and reveals the existence of widespread underproduction in a range of widely-installed software components in Debian.
\end{abstract}


\section{Introduction}

In 2014, it was announced that the OpenSSL cryptography library contained a buffer over-read bug dubbed ``Heartbleed'' that compromised the security of a large portion of secure Internet traffic. The vulnerability resulted in the virtual theft of millions of health records, private government data, and more. OpenSSL provides the cryptographic code protecting a majority of HTTPS web connections, many VPNs, and variety of other Internet services. OpenSSL had been maintained through a ``peer production'' process common in Free/Libre and Open Source Software (FLOSS) where software development work is done by whomever is interested in taking on a particular task. For OpenSSL in early 2014, that had involved only four core developers, all volunteers. OpenSSL was at risk of an event like Heartbleed because it was an extraordinarily important piece of software with very little attention and labor devoted to its upkeep \cite{walden_impact_2020,eghbal_roads_2016}. In this paper, we describe an approach for identifying other important but poorly maintained FLOSS packages.

Over the last three decades, millions of people working in FLOSS communities have created an enormous body of software that has come to serve as digital infrastructure \cite{asay_real_2019}.
FLOSS communities have produced the GNU/Linux operating system, the Apache webserver, widely used development tools, and more \cite{crowston_free/libre_2012}.
In an early and influential practitioner account, Raymond argued that FLOSS would reach high quality through a process he dubbed ``Linus' law''
and defined as ``given a large enough beta-tester and co-developer base, almost every problem will be characterized quickly and the fix obvious to someone'' \cite{raymond_cathedral_1999}.
Benkler coined the term ``peer production'' to describe the method through which many small contributions from large groups of diversely motivated individuals could be integrated together into high quality information goods like software \cite{benkler_coases_2002}.

\begin{figure}
\centering
\begin{knitrout}
\definecolor{shadecolor}{rgb}{0.969, 0.969, 0.969}\color{fgcolor}
\includegraphics[width=0.8\linewidth]{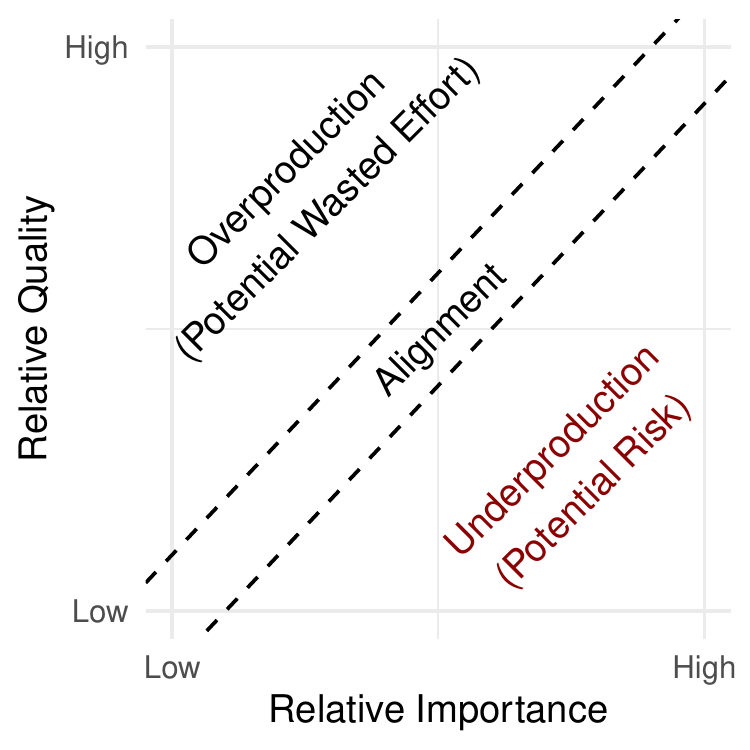} 

\end{knitrout}
\caption{A conceptual diagram locating underproduction in relation to quality and importance. }
\label{fig:simAlign}
\end{figure} 

A growing body of research suggests reasons to be skeptical about Linus' law \cite{schweik_brooks_2008} and the idea that simply opening the door to one's code will attract a crowd of contributors \cite{schweik_internet_2012, benkler_peer_2015}. However, while a substantial portion of labor in many important FLOSS projects is paid \cite{germonprez_rising_2019}, most important FLOSS projects are managed through peer production and continue to rely heavily on volunteer work \cite{eghbal_working_2020}. Many FLOSS projects that incorporate paid labor have limited tools to coordinate or direct work, either paid or volunteer \cite{eghbal_working_2020}. Although some FLOSS projects are now produced entirely within firms using traditional software development models, peer production remains a critical feature of open source software production.

Over time, it has become clear that peer produced FLOSS projects' reliance on volunteer labor and self-selection into tasks has introduced types of risk that traditional software engineering processes have typically not faced. Foremost among these is what we call `underproduction.' We use the term underproduction to refer to the fact that although a large portion of volunteer labor is dedicated to the most widely used open source projects, there are many places where the supply of quality software and volunteer labor is far out of alignment with demand. 
Because underproduction may go unnoticed or unaddressed until it is too late,
we argue that it represents substantial risk to the stability and security of software infrastructure. As a result, we set our key research question as: 
\textit{How can we measure underproduction in FLOSS?}, which we seek to answer both conceptually and empirically. 

Our paper contributes to software engineering research in three distinct ways. First, we describe a broad conceptual framework to identify relative underproduction in peer produced FLOSS repositories: identifying software packages of lower relative quality than one would expect given their relative popularity. Second, we describe an application of this conceptual framework to a dataset of 21,902 source packages from the Debian GNU/Linux distribution using measures derived from multilevel Bayesian regression survival models.
Finally, we present results from two experiments. The first experiment identifies a pool of relatively underproduced software in Debian. The second experiment seeks to validate our application of our framework for identifying underproduction by correlating underproduction with an alternate indicator of risk.

The rest of our paper is structured as follows. We describe prior work on underproduction in §\ref{sec:background} and present our conceptual framework in §\ref{sec:underprod}. We then describe Debian, our empirical setting, in §\ref{sec:debian} and our approach for applying our framework to Debian in §\ref{sec:application}. We present the results of our two experiments in §\ref{sec:experiments}. Finally, we identify significant threats to the validity of our work in §\ref{sec:limitations} and potential implications of the work in §\ref{sec:discussion} before concluding in §\ref{sec:conclusion}.

\section{Background}
\label{sec:background}

\subsection{Detecting and Measuring Software Risk} 
\label{sec:align}

Prevention, detection, and mitigation of risk in software development, maintenance, and deployment are the subject of substantial research interest [e.g., \citenum{natella_assessing_2016, gritzalis_exiting_2018, meidan_measuring_2018}]. Risk detection and management techniques examine overall system safety, develop predictive models to prioritize effort (such as reliability growth models), and seek development techniques to make a project less error-prone and more fault tolerant \cite{bennett_risk_1996}.
One line of work in software quality analysis and risk detection seeks to identify issues by locating ``bad smells'' and anti-patterns. This includes code smells 
\cite{sobrinho_systematic_2018,santos_systematic_2018}, 
as well as architectural smells (ill-considered fundamental design decisions that may trouble the project later) \cite{le_empirical_2018} and community smells (early warning signs of problems in a community). Of particular interest to both software engineering researchers and practitioners are smells that are empirically related to failures \cite{tamburri_exploring_2019}. 

A range of other strategies have been employed to measure risk. Code-level metrics look at complexity, change-prone code, or defect-prone code. Other approaches consider the extent that a codebase takes adequate preventative measures against risk, such as thorough testing \cite{wong_source_2005}. Finally, multi-factor approaches, as in decision-support analysis, take a risk management point of view and incorporate organizational factors, management practices, and areas of potential risk throughout a project's lifecycle \cite{pasha_critical_2018}.

\subsection{Peer Production and FLOSS}
\label{sec:floss}

Free/Libre and Open Source Software (FLOSS) is software released under a license that allows unrestricted use, modification, redistribution, and collaborative improvements \cite{crowston_effective_2004}. 
FLOSS began with the free software movement in the 1980s and its efforts to build the GNU operating system as a replacement for commercial UNIX operating systems \cite{stallman_free_2002}. Over time, free software developers discovered that their free licenses and practices of working openly supported new forms of mass collaboration and bug fixes \cite{crowston_free/libre_2012}.

`Peer production' is a term coined by Yochai Benkler to describe the model of organizing production discovered by FLOSS communities in the early 1990s that involved the mass aggregation of many small contributions from diversely motivated individuals. Benkler \cite{benkler_peer_2015} defines peer production for online groups in terms of four criteria: (1) decentralized goal setting and execution, (2) a diverse range of participant motives, including non-financial ones, (3) non-exclusive approaches to property (e.g. copyleft or permissive licensing), and (4) governance through participation, notions of meritocracy, and charisma, rather than through property or contract. 
In Benkler's foundational account \cite{benkler_coases_2002}, archetypes of peer production include FLOSS projects like the GNU operating system or the Linux kernel as well as efforts like Wikipedia.

\subsection{Systematic comparison of FLOSS in software repositories}
\label{sec:repos}

The process of building and maintaining software is often collaborative and social, including not only code but code comments, commit messages, pull requests, and code reviews, as well as bug reporting, issue discussing, and shared problem-solving \cite{robles_tools_2009}. Non-code trace data may include signals of technical debt \cite{zampetti_automatically_2020}, signs that a given code commit contains bugs \cite{falcao_relating_2020}, or serve as indicators of committed developers, a high-quality software project, or a healthy, sustainable project community \cite{dabbish_social_2012, coelho_identifying_2018, valiev_ecosystem-level_2018}.
Prior research has found that digital trace data capturing online community activity can provide significant insight into the study of software \cite{dabbish_social_2012}. 

These collaborative and social systems offer a data source for understanding both developer team productivity, as in Choudhary's \cite{choudhary_using_2020} study of ``collaboration bursts'' as well as for analyzing macro-level dynamics of software production. For example, Gonzalez-Barahona et al.~\cite{gonzalez-barahona_studying_2014} used the repository of \textit{glibc} as a site to evaluate Lehman's ``laws of software evolution''  including the law of organizational stability which states that work rates on a system are constant. The team found that the laws are frequently not upheld in FLOSS, especially when effort from outside a core team is considered. This work suggests that the effort available to maintain a piece of FLOSS software may increase as it grows in popularity. 

Prior studies have suggested that bug resolution rate is closely associated of a range of important software engineering outcomes, including codebase growth, code quality, release rate, and developer productivity \cite{abou_khalil_longitudinal_2019, kim_how_2006, michlmayr_statistical_2006}. By contrast, lack of maintenance activity as reflected in a FLOSS project's bug tracking system can be considered a sign of failure \cite{coelho_why_2017}. 

\section{Conceptual Framework: Underproduction}
\label{sec:underprod}

Repositories of peer produced FLOSS are susceptible to what we call underproduction---a concept and term that we borrow from several Wikipedia researchers who use the term to describe a dynamic that emerges when volunteers self-select into tasks. 
In particular, we are inspired by a study of Wikipedia by Warncke-Wang et al.~\cite{warncke-wang_misalignment_2015} who build off previous work by Gorbatâi \cite{gorbatai_exploring_2011} to formalize what Warncke-Wang calls the ``perfect alignment hypothesis'' (PAH). The PAH proposes that the most heavily used peer produced information goods (for Warncke-Wang et al., articles in Wikipedia) will be the highest quality, that the least used will be the lowest quality, and so on. 
In other words, the PAH proposes that if we rank peer production products in terms of both quality and importance---for example, in the simple conceptual diagram shown in Figure \ref{fig:simAlign}---the two ranked lists will be perfectly correlated.
In Gorbatâi's terminology, misalignment such that quality is high but demand is low results in `overproduction.'
Peer produced goods are `underproduced' when demand is high but quality is low.

In an economic market, supply and demand are said to be aligned through a price mechanism. Alignment is reached because lower demand decreases prices, which disincentivizes production and returns a market to equilibrium.
Because there is no price signal in Wikipedia to bring consumer demands and producer supply into equilibrium, 
it is unsurprising that Wikipedia deviates substantially from the predictions of the PAH \cite{warncke-wang_misalignment_2015}.
Indeed, ``perfect alignment'' serves not as a serious prediction of the relationship between  Wikipedia articles' quality to the interests of the general public but as a baseline from which to identify lacunae in need of attention \cite{warncke-wang_misalignment_2015}.  Research on Wikipedia has sought to characterize the negative impacts on information consumers from divergence from the PAH baseline and to identify sociological processes through which underproduction might emerge \cite{warncke-wang_misalignment_2015, gorbatai_exploring_2011}.

Despite the central role that FLOSS plays in peer production, we know of no efforts to conceptualize or measure underproduction in software. 
We find this surprising for two reasons. 
First, widespread underproduction seems likely in FLOSS given that FLOSS is characterized by self-selection of software developers into tasks, varying motives among contributors, and the frequent absence of market forces for allocating producers' labor \cite{lakhani_why_2005, crowston_free/libre_2012, oneil_cyberchiefs_2015}. 
Second, the consequences of underproduction are particularly stark in FLOSS where popular software acts as infrastructure \cite{eghbal_roads_2016, eghbal_working_2020}. A low quality Wikipedia article on an extremely popular subject seems likely to pose much less risk to society than a bug like the Heartbleed vulnerability described earlier which could occur when FLOSS is underproduced. 
In this way, underproduction in software reflects an important, if underappreciated, type of risk in FLOSS. 

To answer our research question (\textit{How can we measure underproduction in FLOSS?}) in conceptual terms, our approach to detecting underproduction in software is composed of five steps as follows:

\begin{enumerate}
    \item Assemble a collection of software artifacts that can be consistently measured as described below. These might be software packages, modules, source files, etc.
    \item Identify one or more measures of quality that can be recorded consistently and independently (perhaps repeatedly) across each software artifact in the collection.
    \item Similarly, identify a measure of importance that can be recorded consistently and independently across the collection.
    \item Propose an \textit{ex ante} theoretical baseline relationship between the two measures that reflects alignment. Although this might involve any number of assumptions about an ideal relationship, this might also be a non-parametric claim that the relative ranking of artifacts in terms of quality and importance will be perfectly correlated.
    \item Measure deviation from this theoretical baseline across artifacts.
\end{enumerate}

\noindent The measure of deviation resulting from this process serves as our measure of (mis-)alignment between quality and importance (i.e., over- or underproduction). 
 
In a sense, our conceptual approach involves laying out software packages on dimensions similar to those shown in Figure \ref{fig:simAlign}, empirically identifying an ideal relationship that reflects general alignment (i.e., the zone in the lower left to upper right diagonal), and then measuring deviation from that ideal. This basic conceptual framework can incorporate any number of ways of measuring quality and importance---both areas of active work in software engineering research. Our approach can be carried out using a range of techniques for identifying alignment including entirely non-parametric rank-based approaches, machine learning-based ordinal categorization, or parametric regression-based techniques.

\section{Empirical Setting}
\label{sec:debian}

The first step of applying our conceptual framework involves assembling a collection of software artifacts. We draw our collection from the Debian GNU/Linux distribution which acts as the empirical setting for all of our experiments. GNU/Linux distributions are collections of software that have been integrated, configured, tested, and packaged with an installer. The contributor community producing the distribution focuses primarily on the production of packages and package management tools for managing the installation and updating of software products produced by others. Distributions like Debian play an important role in the FLOSS ecosystem and are the way that the vast majority of GNU/Linux users install operating system software as well as most applications and their dependencies. 
With a community in operation since 1993, Debian is widely used and is the basis for other widely-used distributions like Ubuntu. Debian had more than 1,400 different contributors in 2020\footnote{\url{https://contributors.debian.org/} (Archived: \url{https://web.archive.org/web/20201107231239/https://contributors.debian.org/})} and contains more than 20,000 of the most important and widely used FLOSS packages.

Debian provides detailed and consistently measured longitudinal data on all its packages and maintainers in the form of released databases, datasets, and APIs \cite{caneill_debsources_2016, nussbaum_ultimate_2010, hindle_mining_2010}. 
A body of research in software engineering has used this open data from Debian to understand a range of software development practices. The Debian distribution has served as a basis for applying techniques to detect and mitigate defects \cite{chen_large-scale_2007, michlmayr_statistical_2006}, 
predict bugs and vulnerabilities \cite{pati_comparison_2014},
 detect the evolution of package incompatibilities \cite{claes_historical_2015}, 
predict component reuse \cite{spaeth_lightweight_2008}, demonstrate code clone detection techniques \cite{cordy_debcheck_2011},
develop generalizable QA techniques for complex projects \cite{nussbaum_ultimate_2010},
investigate package dependencies \cite{galindo_duarte_debian_2010}, and as an example of an information processing network \cite{villegas_evolution_2020}. 

\section{Application of Framework}
\label{sec:application}


\subsection{Step 1: Assemble a collection of artifacts}
\label{sec:step1}

Our unit of analysis is the Debian \textit{source package}. Source packages are the format that Debian's package maintainers modify and publish, but they are not used directly by end-users. Instead, source packages are built by computers in a Debian network of ``build daemons'' into one or more \textit{binary packages} that may, or may not, include architecture-specific executables. These binary packages are then distributed to end-users and installed on their computers. Debian also provides tools to allow users to download and build their own binary packages from corresponding source packages. A single source package may produce many binary packages. For examples, although it is an outlier, the Linux kernel source package produces up to 1,246 binary packages from its single source package (most are architecture specific subcollections of kernel modules).

The one-to-many relationship between source and binary packages presented a challenge for our analysis. Although our chosen measure of quality (bug resolution, described in  §\ref{sec:step2}) uses information stored at the level of the source package \cite{davies_perspectives_2010}, our chosen measure of importance (installations, described in  §\ref{sec:step3}), is aggregated at the binary level.
To map source packages to binary packages, we used the Debian snapshot database's public APIs\footnote{\url{https://snapshot.debian.org/} (Archived: \url{https://perma.cc/HQW7-R4Y2})} to identify all binary packages produced by all versions of every source package. 

\subsection{Step 2: Identify a measure of quality}
\label{sec:step2}

The second step of our framework involves identifying a measure of quality for each Debian source package. Quality in software is difficult to measure and common strategies for measuring quality include analyzing bug counts [e.g. \citenum{ray_large_2014}] or assessing code internal design using a series of heuristics [e.g. \citenum{santos_systematic_2018}]. However, software engineering researchers have noted that the quantity of bugs reported against a particular piece of FLOSS may be more related to the number of users of a package \cite{herraiz_impact_2011, davies_perspectives_2010}, or the level of effort being expended on bug-finding \cite{walden_impact_2020} in ways that limit its ability to serve as a clear signal of software quality. In fact, Walden \cite{walden_impact_2020} found that OpenSSL had a lower bug count before Heartbleed than after.
Walden \cite{walden_impact_2020} argued that measures of project activity and process improvements are a more useful sign of community recovery and software quality than bug count. 

Additionally, techniques to assess codebases using code design heuristics are not oriented to the work of a distribution development community. The primary focus of work in a community like Debian is on configuring and testing packages for interoperability, rather than making changes to internal code design. 
In applying our method to Debian for this study, we follow a series of authors who have argued for a focus on a community's effectiveness at resolving the inevitable issues that arise instead of artifact-focused measures \cite{eghbal_working_2020, ronchieri_metrics_2018, adewumi_systematic_2016, crowston_free/libre_2012, ruiz_measuring_2011, aksulu_comprehensive_2010}. 
Specifically, our measure of quality is the speed at which bugs are resolved. We treat the difference between opening and closing times as the \textit{time to resolution}. Time to resolution has been cited as an important measure of FLOSS quality by a series of software engineering scholars \cite{eghbal_working_2020, abou_khalil_longitudinal_2019, crowston_free/libre_2012, kim_how_2006}. Although there may be many reasons for protracted resolution time in a distribution (e.g. maintainer skill, task complexity, report quality, lack of resources, bugs in the underlying software package causing packaging problems), the unresolved bug still represents an issue for the distribution's maintainer community and a problem for end users, whatever the reason.

Calculating this measure requires interpretation of bug reports and resolution data which Debian tracks using a database where each transaction takes the form of a specially formatted e-mail message. 
We extract comprehensive data on bugs from the Debian Bug Tracking System which is also referred to as ``debbugs'' or the ``BTS.'' After obtaining an archival copy of all bugs from the BTS, we queried the Ultimate Debian Database \cite{nussbaum_ultimate_2010} to map bugs to packages. We parsed all actions (i.e., e-mail messages) associated with each bug into columnar data for analysis.
Using BTS data, we identified the date and time when each bug was opened and closed. We treated the marking of a bug as ``closed,'' ``forwarded'' (i.e., not solved by the maintainer but rather referred to the ``upstream'' development team for the package itself and therefore no longer Debian's responsibility), or ``merged'' (i.e., designated as a repeat report of an issue already in the database) as closed.

Our approach differs from the approach taken by Zerouali et al.~\cite{zerouali_relation_2019} who measure bugs as closed based on when they are last modified. We diverge from their method because final modification to a bug typically occurs when a bug is archived through an automated process that occurs about 30 days after closure. In our examinations of the database, we found that this process can be inconsistent and that unarchiving can occur as a function of administrative processing as well as due to true reopening of a bug. 

\subsubsection*{Controlling for Bug Severity}

A limitation relating to our measure of quality relates to the fact that not all bugs require equal attention. In recognition of this fact, bugs in Debian are assigned a \textit{severity} by the submitter which can be modified by the package maintainer or others.  Severity in Debian is one of the following categories: wishlist, minor, normal, important, serious, grave, and critical (in that order) with normal as the default.
We extracted severity for every bug from the BTS to use as a control. We treat the  4,559 bugs where the severity is ``not set'' and the 2,239 bugs where the severity is listed as ``fixed'' as priority ``normal.''\footnote{The ``fixed'' severity was used from 2000 to 2002 to indicate a non-maintainer upload and then deprecated in favor of a ``tagging'' approach for this type of fix.}
We omit all ``wishlist'' severity bugs (124,961) because in addition to being used to track bugs that are ``very difficult to fix due to major design considerations,'' this category may be used for a wide range of non-bug ``to-do list'' items. 

Although there may be many other reasons for extended resolution time (e.g. skill, resources, inherent difficulty of the task), we do not include controls for these considerations. This is because our goal is to capture variation in resolution rate due to a range of reasons, which themselves may correspond to a level of risk. As a result, we at correspond to levels of risk. as a result we stratify on. describe threat and what we do to address the threat

\begin{figure}
\includegraphics[width=\linewidth]{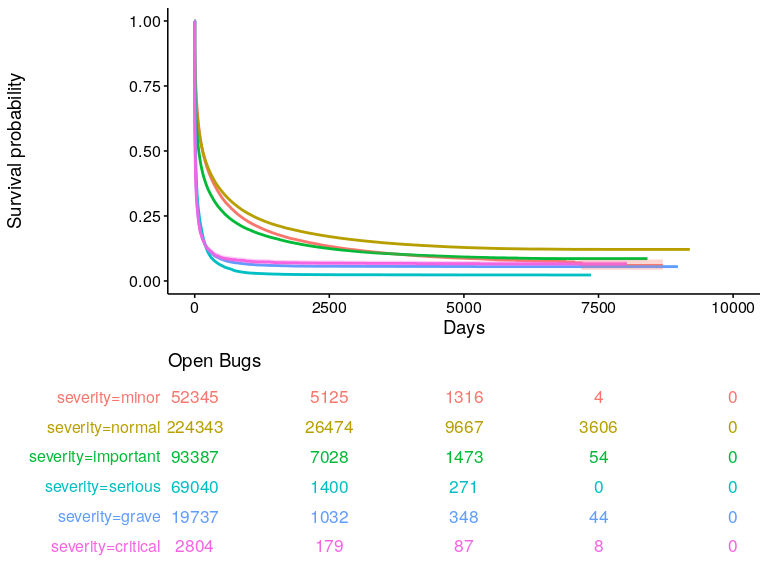}
\caption{A Kaplan-Meier curve that shows the number of bugs of different severities that remain open over time. \label{fig:kmSurvival}}
\end{figure} 

Figure \ref{fig:kmSurvival} shows non-parametric Kaplan-Meier survival curves for bugs of differing severities. These curves depict the probability of a given bug going from unresolved to resolved in days since it was filed.
We observe in this plot that the curves are tightly clustered during the first few days after a bug is reported: about half of all bugs, regardless of severity, are solved shortly after they are reported.
Bugs that are serious, grave, or critical are considered ``release-critical'' and must be fixed for the next ``stable'' release of Debian. Release-critical bugs include security problems, bugs that make packages unusable, or licensing requirements that are incompatible with Debian.\footnote{\url{https://www.debian.org/Bugs/Developer} (Archived: \url{https://perma.cc/MX3T-36PP})} 
We observe that the curves in Figure \ref{fig:kmSurvival} cluster into two general shapes that correspond to release critical and non-release critical bugs, and that release-critical bugs are fixed more quickly. 

\subsubsection*{Modeling \textit{time to resolution} using Bayesian survival models}

Measuring quality as \textit{time to resolution} poses two analytic challenges. The first challenge is the fact that a substantial portion of bugs remain unresolved at the point of data collection. Because bugs languishing in an open state is precisely the type of risk we hope to measure, omitting bugs that are open at the time of data collection would underestimate how quickly bugs are resolved. Following recent work in software engineering \cite{abou_khalil_longitudinal_2019}, we incorporate data on unresolved bugs by modeling the process of bug closure directly using Cox proportional hazard models.\footnote{Cox models are a type of survival model developed originally in epidemiology to estimate how a behavior or treatment might prolong or shorten patients' lives while incorporating data from individuals for whom data is censored (i.e., individuals who are still alive at the conclusion of data collection). In our case, data is censored for any open bug.} 

A second challenge in measuring quality as \textit{time to resolution} comes from the fact that the distribution of bugs across packages is highly unequal. Most of the packages we examine (14,604 of 21,902) have 10 or fewer bugs and more than one out of six (3,857 of 21,902) have only one bug reported.
In response, we use Bayesian hierarchical models fit using Stan. The intuition behind this choice is that when a package has few bugs, the overall problem resolution rate across Debian is more informative than the package's small number of data points. In general, the Bayesian approach allows us to become progressively more confident in an estimate when more bugs have been filed against a package. 

Our approach finds support in the argument made by Ernst \cite{ernst_bayesian_2018} who elaborates the value of Bayesian hierarchical modeling as a method for carefully distinguishing local and global characteristics in studies of software repositories.
Our approach offers a ``partial-pooling'' model that allows us to both take advantage of the structure of our data (bugs are clustered within packages) and to update our prior assumption about expected resolution rates based on the new information that each bug supplies.


We incorporate our control for \textit{severity} into a survival model of the following form where bugs are the unit of analysis and where $\lambda$ reflects the hazard function  capturing whether bug $k$ in package $j$ will be resolved at time $t$. We represent
the severity for each bug as $x_{jk}$  and use $z_j$ to reflect the log of the package-level random effect for package $j$.
\footnote{We use notation for random effects survival models drawn from Sargent \cite{sargent_general_1998}.}

\begin{equation*}
\lambda (t; x_{jk}; q_j) = \lambda_0 (t) exp(\beta x_{jk} + q_j)
\end{equation*}

\noindent Our measure of quality ($q_j$) is the package-level random effect of the posterior distributions in our survival models. We estimate this quantity using 4,000 independent draws from each package's posterior distribution using Stan. We take 95\% credible intervals of these empirical distributions to reflect uncertainty.  Estimates of $q_j$ effects for all 21,902 packages are reported in our supplement described in §\ref{sec:supplement}.

\subsection{Step 3: Identify a measure of importance}
\label{sec:step3}

Our third step involves identifying a consistent measure of importance for every artifact in our collection.
In FLOSS contexts, importance can be measured as attention on hosting sites, number of active users, intensity of use, dependency networks, or criticality of function (e.g., life-safety systems) \cite{boldyreff_heartbeat:_2009, eghbal_working_2020, stergiopoulos_time-based_2016}.
We measure importance using data from from the Debian ``Popularity Contest'' or ``Popcon'' application. Popcon is an opt-in survey that shares anonymous data from volunteer systems back with Debian.  Popcon data has been used in a range of previous studies in software research \cite{boldyreff_heartbeat:_2009, herraiz_impact_2011, davies_perspectives_2010}. 
Popcon is particularly applicable in our case because it is a signal which Debian itself has developed and deployed and because it is displayed in multiple locations in Debian's maintenance platforms. 
One important limitation of Popcon data is that despite the millions of systems running Debian, only a fraction have opted to report installation data. 

Our study includes data from 201,484 systems from a single-day snapshot on July 6, 2020.  Popcon includes two measures: \textit{inst} for \textit{installation} (i.e. the presence of a package on a machine), which serves as our measure of importance for subsequent analysis ($i_j$, for each package $j$); 
and a measure called ``vote'' which attempts to capture if packages are being used. In our analysis, we use \textit{installation}; our supplement  §\ref{sec:supplement} includes a version of our analysis conducted using ``vote'' as our measure of importance. 

Unfortunately, because Popcon reports installation at the binary level, we cannot use this data to distinguish whether a single individual reported the installation of multiple binaries associated with a given source package. To avoid double-counting, we use the binary-source mappings described in §\ref{sec:step1} to set $i_j$ to the largest install count among all binary packages associated with a source package $j$. As a result of this construction, our installation measure is necessarily conservative but we can be assured that a source package was installed at least $i_j$ times.

\subsection{Step 4: Select a baseline relationship}
\label{sec:step4}

The fourth step in our conceptual framework involves comparing measures of quality and importance to some ideal baseline relationship. For our baseline, we take a non-parametric ranking approach that is similar to, but more granular than, the approach taken in Warncke-Wang et al.'s Wikipedia analysis which uses Wikipedia quality categories \cite{warncke-wang_misalignment_2015}. Our baseline definition of alignment is when the relative rankings of importance and quality are the same ($rq_i = rq_j$). We treat a ranking of 1 as describing the worst observed quality or lowest number of installations, while a rank of 21,902 represents the highest.

\subsection{Step 5: Measure deviation}
\label{sec:step5}

The final step of our conceptual framework involves measuring deviation from our theoretical baseline. We describe our measure of alignment as the ``underproduction factor'' ($U_j$) which we measure as the log of the ratio of rankings of importance and quality:

\begin{equation*}
U_{j} = \log{\frac{ri_j}{rq_{j}}}
\end{equation*}

\noindent Given this construction, $U_{j}$ will be zero when a package is fully aligned, negative if it is overproduced, and positive if it is underproduced. This approach means that the range of $U_j$ is a function of repository size. In our dataset of 21,902 source packages, the range of $U_j = [-10, 10]$, where $U_j=10$ for a theoretical maximally underproduced package where $ri_j = 21,902$ and $rq_j = 1$ .

Although constructing $U_j$ for a single value of $q_j$ is straightforward, incorporating uncertainty in our measure of quality requires additional work. 
Because posterior draws in Stan are independent, we can incorporate uncertainty in our measure of quality by computing $U_{j}$ using the quality ranking from each of 4,000 posterior draws taken from the estimated random effect for each package $j$.
$U_j$ reported in our analysis reflect the 95\% credible intervals from $U_j$ computed separately for these draws. Because we only have a single measure of installation $i_j$, we do not attempt to incorporate uncertainty in this measure into our analysis.

\section{Results from Experiments}
\label{sec:experiments}

In order to assess our conceptual model and to provide an empirical answer to our research question,
we conduct two experiments. 
Our first experiment describes results from the application of our method described in §\ref{sec:application} and suggests that a minimum of 4,327
packages in Debian are underproduced. Our second experiment validates our approach using an alternate measure of risk.

\subsection{Experiment 1: Identifying Underproduced Software}
\label{sec:debian_experiment}

\begin{figure}
\centering
\includegraphics[width=\linewidth]{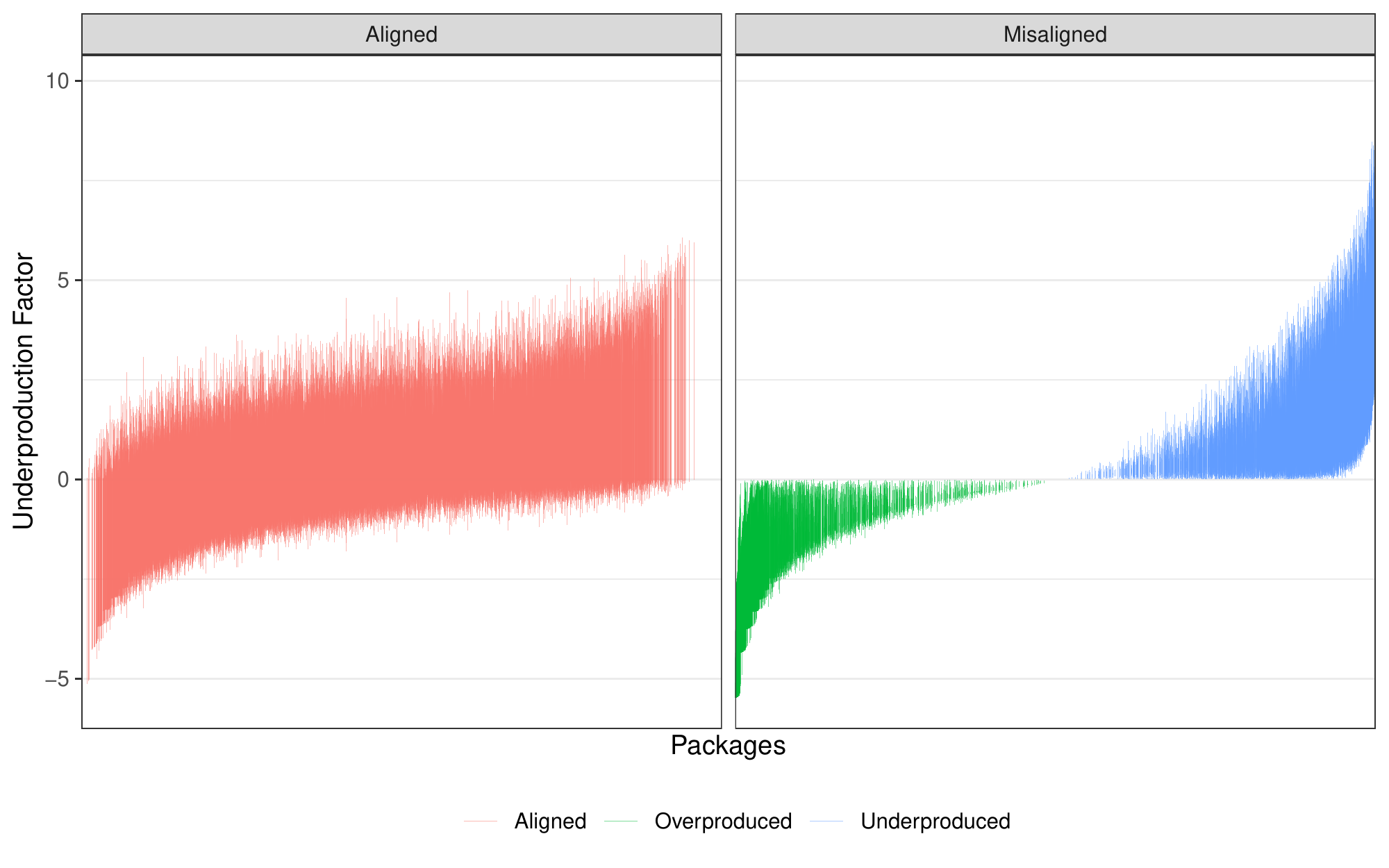}
\caption{Credible intervals  for $U_j$ for every package in Debian. We treat all packages whose CIs include zero as ``aligned''; those whose CIs are entirely above 0 are labeled ``underproduced;'' those whose CIs are entirely below zero are labeled ``overproduced.'' \label{fig:caterpillar}}
\end{figure} 

\begin{figure}
\includegraphics[width=\linewidth]{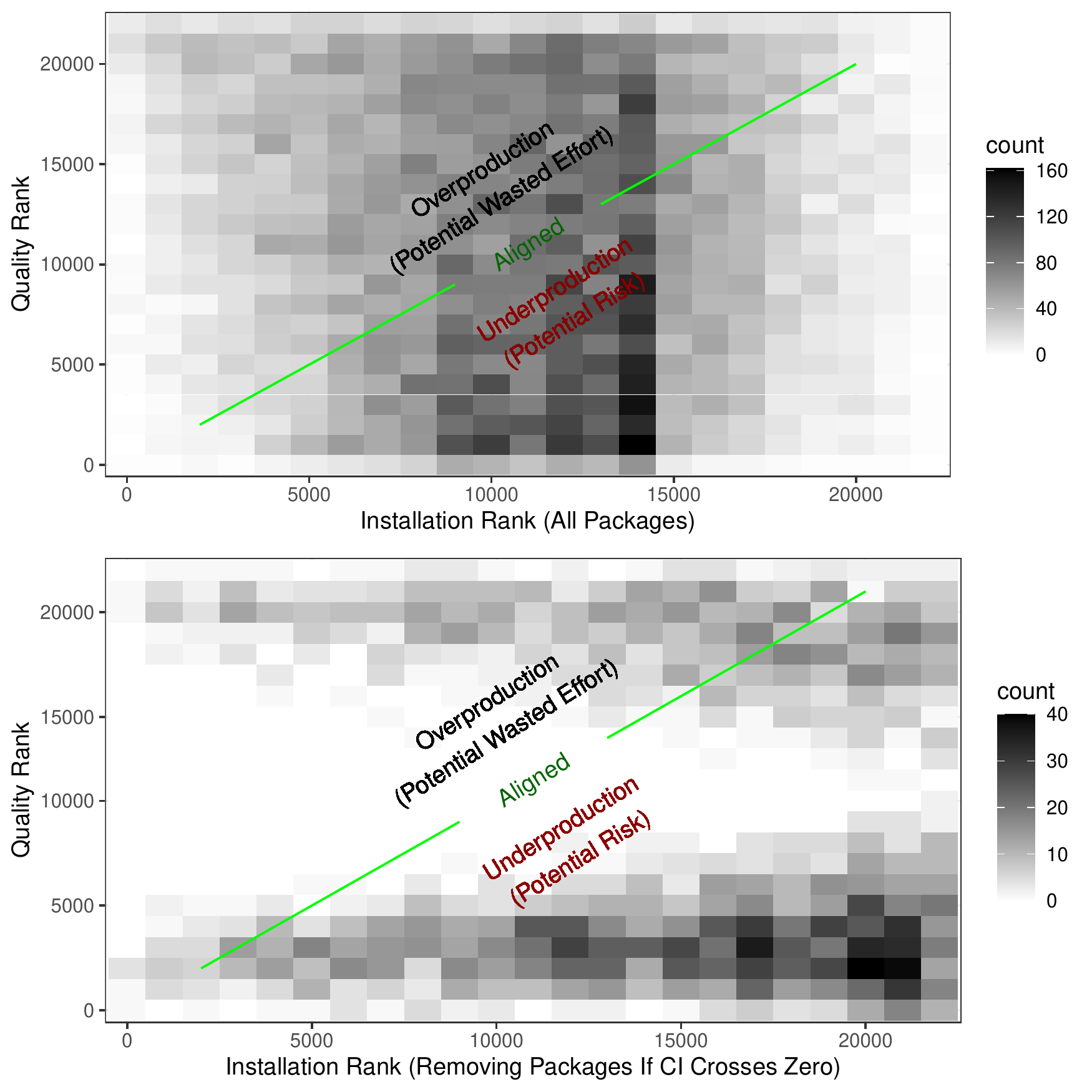}
\caption{A heatmap of software alignment. Color intensity indicates the number of packages occupying a given ranking of quality and installation. Aligned packages appear along the lower-left to upper-right diagonal. The top heatmap includes all packages, while the bottom heatmap contains only those packages for which the 95\% credible interval does not cross zero.\label{fig:heatmap_both}}
\end{figure} 

Our approach provides evidence that underproduction is widespread in Debian. Figure \ref{fig:caterpillar} shows 95\% credible intervals (CIs) for all 21,902 packages. We describe packages whose 95\% credible interval for $U_j$ includes zero ($0 \in U_j$) as ``aligned'' and packages where both ends of the credible interval have the same sign as ``overproduced'' ($U_j<0$) and ``underproduced'' ($U_j>0$). The wide credible intervals for many of the packages shown on the left panel of Figure \ref{fig:caterpillar} reflect the fact that many of the aligned packages may be misaligned packages with high variance. The noise in our measures of $q_j$ and $U_j$ is partially attributable to the fact that many packages have few bugs.

Figure \ref{fig:heatmap_both} displays a heatmap visualization that shows the number of packages occupying a range of installation ranks ($ri_j$) along the $x$-axis and quality ranks ($rq_j$) along the $y$-axis in evenly spaced bins. In this way, this non-parametric data visualization seeks to reflect the basic intuition that goes into the construction of our measure of underproduction $U_j$.
Because $rq_j$ is a distribution, the figures display a ranking of the mean value of $q_j$. 
The top panel shows all packages in Debian and clearly shows data spread across the heatmap rather than clustered along the diagonal. The relative lack of density along diagonal in the top figure is evidence that misalignment in Debian is widespread. 
The relatively high density of packages in the lower right indicates that underproduction in Debian is common. 

\begin{figure}
\includegraphics[width=\linewidth]{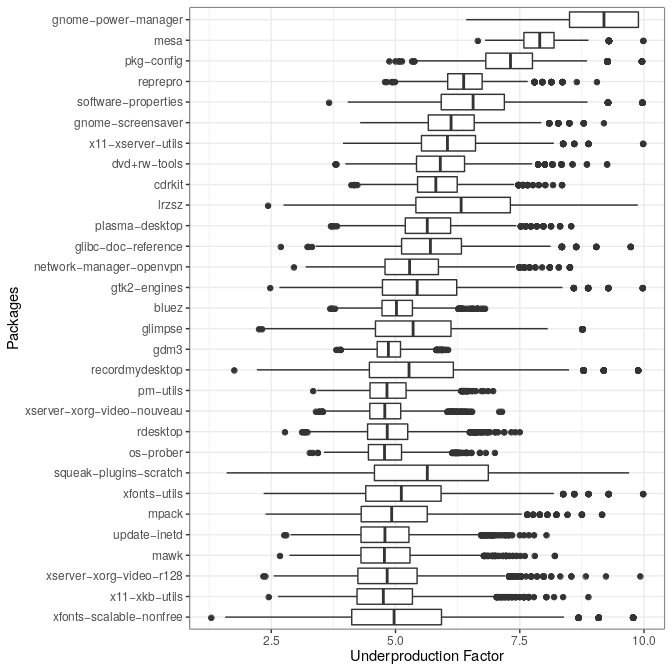}
\caption{Packages displaying the highest mean levels of underproduction. Boxplots show the mean and interquartile range of our distributions of $U_j$ and reflect uncertainty in our model of package-level quality.}
 \label{fig:worstOfWorst}
\end{figure} 
The lower panel of Figure \ref{fig:heatmap_both} shows only packages for which the 95\% credible interval excludes zero (i.e., the misaligned packages shown on the right panel of Figure \ref{fig:caterpillar}). Removing aligned packages reveals much more density among underproduced packages relative to overproduced packages. Removing `aligned' packages visibly hollows out the center of the heatmap filled with packages of moderate quality because many of these packages have wide CIs. However, although there are almost no packages of middling quality that we can confidently say are overproduced, there are many that we can confidently say are underproduced. Further, there is substantial density---hundreds of packages---in the extreme bottom right corner. This area contains packages that are almost maximally underproduced according to our measure.
In Figure \ref{fig:worstOfWorst}, we present a list of the packages displaying the highest levels of underproduction (mean $U_{j}$) with per-package boxplots drawn from the posterior distribution associated with each package. 

\subsection{Experiment 2: Validation Using Alternate Indicator}
\label{sec:validation}

To validate our application of our conceptual framework, we test whether our measure of underproduction can be used to predict community responses to risk.
To do so, we collect data on the count of ``non-maintainer uploads'' (NMUs) which occur when any individual other than the denoted maintainer updates a version of a package in Debian.\footnote{\url{https://wiki.debian.org/NonMaintainerUpload} (Archived: \url{https://perma.cc/TYN9-WQ79})} 
Although some package maintainers may welcome NMUs, or even invite them, NMUs are widely understood as an indicator of risk in Debian. One of the criteria for ``package salvage,'' that is, the taking over of maintainership without cooperation from the designated maintainer, is the presence of NMUs.\footnote{\url{https://wiki.debian.org/PackageSalvaging} (Archived: \url{https://perma.cc/NBF9-6QSK})} We hypothesize that any valid measure of underproduction in Debian will be positively associated with the number of NMUs that the package receives.


\begin{table}
\begin{center}
\caption{Predicting non-maintainer upload (NMU). \label{tab:validityModels}}
 \begin{tabular}{l c} \hline Intercept      & $-3.03^{*}$       \\                & $ [-3.10; -2.95]$ \\ Mean $U_j$     & $0.47^{*}$        \\                & $ [ 0.40;  0.53]$ \\ \hline Num. obs.      & $21902$           \\ \hline \multicolumn{2}{l}{\scriptsize{$^*$ 0 outside the confidence interval.}} \end{tabular}
\end{center}
\end{table}

Given that our measure of NMUs is an overdispersed count, we conduct our analysis using a negative binomial regression framework. 
Results from our model are reported in Table \ref{tab:validityModels}.  
We find that $U_j$ (underproduction factor) is a statistically significant predictor of NMU count, and that increases in underproduction factor correlate with increased numbers of NMUs.  
Our model estimates suggest that there is a strong relationship between our measure of underproduction and NMU count ($\beta=0.47$; $\mathrm{CI}=[0.40, 0.53]$).
To illustrate this effect, consider two prototypical packages: a fully aligned package ($U_j=0$) and one of the most underproduced packages in the distribution as depicted in Figure \ref{fig:worstOfWorst} ($U_j=5$). Our model predicts that, on average, a prototypical aligned package will be NMUed extremely rarely ($\widehat{\mathrm{NMU}}=0.048$; or less than 5\% of packages will have a single NMU). On the other hand, it suggests that our prototypical underproduced package will be NMUed as often as not ($\widehat{\mathrm{NMU}}=0.504$).

\section{Threats to Validity}
\label{sec:limitations}

Our experimental findings may be limited in their generalizability. Although Debian is widely used in software engineering research and the study of distributions has been identified as preferable to code repository hosting platforms 
\cite{spaeth_sampling_2007}, it is only a single empirical setting.
Additionally, Debian is idiosyncratic in ways that might threaten our ability to generalize. For example, Debian's policy to only include FLOSS means we do not assess widely-installed packages with nonfree licenses or any tools that are not packaged in Debian.
Additionally, although our sample captures a broad timespan of bug reports in Debian, it only reflects one snapshot in time for package installation. 
As a result, packages may have changed in importance over time in ways that our analysis does not capture.
We also cannot know how our experimental findings might generalize to smaller, newer, more commercial, or more narrowly focused development communities. 

With respect to internal validity, it is important to note that our results are only correlational. While prior research has implicated lack of maintenance as increasing the likelihood of failure, 
we do not develop evidence to support a causal claim 
[e.g. \citenum{walden_impact_2020,coelho_identifying_2018,eghbal_roads_2016}]. We hope that further work will demonstrate whether underproduction is predictive of significant failure. The validity of our approach is also subject to threats related to construct validity. Underproduction is a concept borrowed from economics and involves a relationship between supply and demand. Although we have leveraged existing studies of underproduction in Wikipedia as part of our process of conceptualization, there remains space for further discussion about what should constitute `supply' and `demand' in FLOSS. For example, we treat supply as quality but it might also be conceptualized as code quantity or developer effort. Should demand be a raw measure of consumption, as we have conceptualized it, or should we explore alternate approaches like central positions in software dependency networks? 

In a related sense, our empirical work is subject to threats that stem from choices we made in operationalization. For example, resolution time is an imperfect and partial measure of quality. Although we have omitted bugs that package maintainers reject, certain packages might receive higher numbers of low quality or difficult-to-reproduce bug reports from less technical users, prolonging resolution time. The very high bar to filing a bug in Debian may mitigate this concern.\footnote{The onerous process of filing a bug report in Debian is described here: \url{https://www.debian.org/Bugs/Reporting} (Archived: \url{https://perma.cc/RG3U-ZC7J}).} Additionally, resolution time is limited in that it emphasizes the quality-in-use perspective rather than directly taking up artifact concerns (e.g., is the code written in ways that make it efficient to maintain). Applying artifact-based metrics from software engineering to GNU/Linux distributions like Debian remains an active area of research in software engineering. We welcome future attempts to integrate alternate metrics into our framework.

Our measure of installation is constrained to the systems whose administrators have volunteered data using Popcon. As with all opt-in surveys, this results in a non-random sample. Bias in this sample is possible because the installation of certain packages is possibly correlated with participation in the survey.
It is also the case that importance and install base may be measured in different ways and our choice of metric may influence our results \cite{zerouali_diversity_2019}.
Although we consulted with Debian community members in designing and interpreting our experiments, there is no ground truth that we can use to ensure that we got it right. 

Our application of our approach to underproduction analysis is limited in that 
the \textit{ex ante} baseline we selected to demonstrate our approach relies on rank ordering and can only identify \textit{relative} underproduction within a group of software components. 
If our method were applied to a collection of software components where all software was underproduced, it would be able to identify the worst of the batch but could not reveal a high degree of risk in general.
Although this decision side-steps the need for parametric assumptions, an ordered ranking also means that we do not distinguish between consecutively ranked components.

Finally, we validated our approach using non-maintainer upload (NMU) count: the number of times a package is updated by someone other than its maintainer. However, this measure is not an entirely independent measure: packages with long resolution times and high user bases may also be more likely to draw outside contributions in the form of NMUs. 

\section{Discussion}
\label{sec:discussion}

Our results suggest that underproduction is extremely widespread in Debian.
Our non-parametric survival analysis shown in Figure \ref{fig:kmSurvival} suggests that Debian resolves most bugs quickly and that release-critical bugs in Debian are fixed much more quickly than non-release-critical bugs. The presence of substantial underproduction in widely-installed components of Debian 
exposes Debian's users to risk.  We explore several implications of these findings in the sections below.

\subsection{The Long Tail of GUI Underproduction}

One striking feature of our results is the predominance of visual and desktop-oriented components among the most underproduced packages (see Figure \ref{fig:worstOfWorst}). Of the 30 most underproduced packages in Debian, 12 are directly part of the XWindows, GNOME, or KDE desktop windowing systems. For example, the ``worst'' ranking package, GNOME Power Manager (\textit{gnome-power-manager}) tracks power usage statistics, allows configuration of power preferences, screenlocking, screensavers, and alerts users to power events such as an unplugged AC adaptor. Seven additional packages in Figure \ref{fig:worstOfWorst} are also oriented to desktop uses. For example, the \textit{pm-utils} will suspend or hibernate a computer, \textit{network-manager-openvpn} manages network connectivity through VPNs, Ethernet, and WiFi, \textit{mesa} is a 3D graphics library, \textit{bluez} is part of the Linux Bluetooth stack, \textit{cdrkit} and \textit{dvd+rw-tools} are both tools for creating CDs and DVDs, and \textit{recordmydesktop} is a screen capture program. Our finding of relative underproduction in these programs appears to be in line with the history of critiques of GNU/Linux with respect to desktop usability and visual tools \cite{boldyreff_survey_2009}. 
Although some of the reported issues in these GUI tools may be aesthetic,
inspection reveals flaws reported at a range of severities including many very serious bugs. 

A critique of the significance of this result might be that visual components are less likely to be used in business-critical circumstances. However, these packages have enormous install bases and are relied upon by many other packages.
These results might simply reflect the difficulty of maintaining desktop-related packages. For example, maintaining \textit{gnome-power-manager} includes complex integration work that spans from a wide range of low-level kernel features to high-level user-facing and usability issues. This pattern of underproduced GUI components suggests that although desktop GNU/Linux has made substantial progress, it remains a source of risk.

\subsection{Implications for Software Engineering Research}

Although our conceptual model and experiments demonstrate that underproduction can be measured, the detection and measurement of underproduction and the modeling of its predictors, causes, and remedies reflect a series of open challenges for the software engineering research community.
More work is needed to further validate our conceptual framework and our statistical approach. We also hope that future work will extend our approach and implement our framework in other repositories of FLOSS.

\subsection{Implications for Practice}

FLOSS communities may find it useful to employ our technique to identify underproduced software in their repositories and to allocate resources and developer attention accordingly. For example, the Debian project has a volunteer QA team who might benefit from using our analysis to allocate its effort.
Although underproduction is likely to be a particularly acute problem in FLOSS projects due to developer self-selection into tasks, it may also exist in non-FLOSS contexts. We look forward to working with managers of software repositories in both FLOSS and non-FLOSS contexts to help them implement and take action based on measures of underproduction.

While FLOSS practitioners may be particularly worried about underproduction in their projects, the software infrastructure risk that results from underproduction in FLOSS is of broader concern. FLOSS acts as global digital infrastructure. Failures in that infrastructure ripple through supply chains and across sectors. Technology leaders may see opportunities to improve their own risk profiles by offering support to FLOSS components identified as relatively underproduced through the type of analysis we have described. 

Finally, many studies have examined the effect of funding when firms participate in FLOSS. Mixed and often ineffective results might be improved if underproduction is taken into account when directing resources to FLOSS projects. Tradeoffs associated with the potentially demotivating effect of money may be mitigated if funds are targeted at high-impact areas with little existing volunteer labor.
The influx of successful investment that followed the Heartbleed vulnerability \cite{walden_impact_2020} may offer concerned organizations some hope that intervention in response to underproduction is possible and can be effective. Our work seeks to help guide these types of interventions before events like Heartbleed.

\section{Conclusion}
\label{sec:conclusion}
Our work makes three important contributions to software engineering research: we present a broad conceptual framework for identifying relative underproduction; we illustrate our approach using data from Debian; and we validate our method using a measure of response to risk.
Results from our experiments revealed significant underproduction in the widely used Debian distribution and suggest that many of the most underproduced packages in Debian are desktop applications. 

Flaws in widely used software components, regardless of their purpose, represent a source of risk to our shared digital infrastructure. Even if a given bug does not result in system failure, it may provide an attack surface for intrusion or block upgrades of other vulnerable or failure-prone components. Despite widespread dependence on FLOSS, the burden of maintenance continues to fall on small teams of volunteers selecting their own tasks. Without fresh investment of skilled and engaged participants, this public resource will remain at risk. As with Heartbleed, underproduction may not be recognized until it is too late. We hope that our work offers a step toward preventing these failures. 

\section*{Online Supplement}
\label{sec:supplement}

Data, code, and supplemental information for this paper are available in the Harvard Dataverse at the following URL: \url{https://doi.org/10.7910/DVN/PUCD2P}

\section*{Acknowledgement}
 The authors gratefully acknowledge support from the Sloan Foundation through the Ford/Sloan Digital Infrastructure Initiative, Sloan Award 2018-11356. Wm Salt Hale of the Community Data Science Collective and Debian Developers Paul Wise and Don Armstrong provided valuable assistance in accessing and interpreting data. Rene Just at the University of Washington generously provided valuable insight and feedback. We are also grateful to our anonymous reviewers who pointed out opportunities for improvement. This work was conducted using the Hyak supercomputer at the University of Washington as well as research computing resources at Northwestern University.

\balance
\bibliographystyle{IEEEtran}
\bibliography{IEEEabrv,bibliography.bib}
   
\end{document}